\documentclass[12pt]{article}
\def\D{\Delta}
\def\d{\delta}

\def\l{\lambda}
\def\S{\Sigma}

\def\g{\gamma}
\def\e{\epsilon}
\def\s{\sigma}

\def\a{\alpha}

\def\m{\mu}
\def\n{\nu}
\def\r{\rho}
\def\s{\sigma}

\def\F{\Phi}
\def\th{\theta}
\def\e{\epsilon}

\def\det{\textrm{det}}
\def\const{\textrm{const.}}
\newcommand{\be}{\begin{equation}}
\newcommand{\ee}{\end{equation}}
\newcommand{\bea}{\begin{eqnarray}}
\newcommand{\eea}{\end{eqnarray}}

\begin{document}

\begin{center}
\bf{SPIN NETWORK WAVEFUNCTION AND THE GRAVITON PROPAGATOR}
\end{center}

\bigskip
\bigskip
\begin{center}
A. MIKOVI\'C
\end{center}

\begin{center} Departamento de Matem\'atica  \\
Universidade Lus\'ofona de Humanidades e Tecnologias\\
Av. do Campo Grande, 376, 1749-024 Lisboa, Portugal\\
and\\
Grupo de Fisica Matem\'atica da Universidade de Lisboa\\
Av. Prof. Gama Pinto, 2, 1649-003 Lisboa, Portugal\\
\end{center}

\centerline{E-mail: amikovic@ulusofona.pt}

\bigskip
\bigskip
\begin{quotation}
\noindent\small{We show that if the flat-spacetime wavefunction in the spin network basis of Loop Quantum Gravity has a large-spin asymptotics given by Rovelli's ansatz then the corresponding graviton propagator has the correct large-distance asymptotics nonperturbatively and independently of the spin foam model used to describe the evolution operator. We also argue that even in the Rovelli approach the wavefunction should satisfy the Hamiltonian constraint and we give an explanation for the spin parameter appearing in Rovelli's ansatz.}\end{quotation}

\bigskip
\bigskip
\noindent{\bf{1. Introduction}}

\bigskip
\noindent The Loop Quantum Gravity (LQG) and the related spin foam approach, see \cite{lqg}, provide quantization procedures for General Relativity which are manifestly independent from a spacetime background metric and hence are nonperturbative in the metric. However, because these are connection based formulations, it is difficult to see what is the effective semiclassical theory in terms of the spacetime metric \cite{mik2}. Rovelli has proposed recently a promissing approach to the problem of semiclassical limit of LQG \cite{ro2,ro3} and the idea is to calculate the graviton propagator within the LQG formalism and to study the semiclassical limit by analysing the large-distance asymptotics of the propagator. In LQG the spacetime geometry is encoded in the spin networks, and large distances correspond to the large-spin limit of the spin network wavefunction. By making a natural assumption that the flat-spacetime wavefunction in the spin network basis has a Gaussian form and by assuming that the time evolution is determined by the Barret-Crane (BC) Euclidean spin foam model, Rovelli was able to show that the LQG graviton propagator has the correct classical large-distance asymptotics \cite{ro2}.

This is a very encouriging result for the problem of semiclassical limit of LQG, and in order to extend it one would like to see how this result depends on the definitions and the assumptions which were used. Note that the definition of a flat-spacetime quantum field theory (QFT) propagator has to be extended to the case of LQG where there is no background metric. Rovelli has defined the LQG propagator by using a non-standard form of the flat-spacetime propagator. In this paper we are going to analyze the Rovelli definition of the LQG propagator and compare it to a more natural definition of the LQG propagator which is based on the standard form of the flat-spacetime propagator. 

This new definition of the LQG propagator allows one to explore the limit which corresponds
to the case of equal-time field correlations in QFT. In this limit the spin foam transition amplitude is an identity matrix and therefore the propagator is solely determined by the spin network wavefunction. Hence if the wavefunction is known, one can obtain a nonperturbative propagator, as oposed to the Rovelli definition, where a nonperturbative propagator can be obtained only if a nonperturbative spin foam transition amplitude is known. 

However, in order to do this one would need a vacuum wavefunction which is a solution of the Hamiltonian constraint. Such a solution has been constructed in the Euclidean LQG case \cite{mik2}. The wavefunction coefficients are given in the spin network basis as three-dimensional state sums of quantum spin network evaluations at a root of unity. Since at the moment there is no efficient technique for calculating these sums, one can try to explore which large-spin asymptotics of the vacuum wavefunction can give the correct classical graviton propagator. A natural large-spin asymptotics is the one given by the Rovelli's ansatz for the wavefunction, and this case will be explored in this paper.

In section 2 we give a LQG definition of the graviton propagator which is a straightforward extension of the QFT definition. In section 3 we derive the Rovelli propagator as a preparation for doing calculations with the new definition. In section 4 we use the new defintion to show that the LQG graviton propagator has the correct semi-classical asymptotics even non-perturbatively if the large-spin asymptotics of the spin network function is given by the Rovelli ansatz. In section 5 we present our conclusions.

\bigskip
\bigskip
\noindent{\bf{2. Graviton propagator in LQG}}

\bigskip
\noindent When the spacetime curvature is small compared to $l_P^{-2}$, where $l_P$ is the Planck length, one can write the spacetime metric as 
\be g_{\m\n}(x) = \eta_{\m\n} + l_P h_{\m\n} (x)\ee
where $\eta = diag(-1,1,1,1)$, $x=x^\m$ denotes the spacetime coordinates and  $h$ is a spin-two field. The corresponding propagator is defined as
\be G_{\m\n,\r\s}(x,y)=\langle 0|T(h_{\m\n}(x)h_{\r\s}(y))|0\rangle \,,\label{prop}\ee
where $|0\rangle$ is the flat spacetime vacuum. In the limit when $\hbar \to 0$ and the Newton constant $G_N \to 0$ one has
\be G(x,y) \approx {\textrm{const.}\over |x-y|^2} \,,\label{sda}\ee
where we have supressed the spacetime indices. This is also the asymptotics for large distances, i.e. when $|x-y|>> l_P$.

In order to extend the definition (\ref{prop}) to the case of LQG, let us write
the spatial components of (\ref{prop}) as
\be G_{mn,rs}(t,\vec x;t',\vec y)=\langle 0|\,h_{mn}(t,\vec x)h_{rs}(t',\vec y)|0\rangle \,,\label{qftp}\ee
where $x^\m =(t,x^m )=(t,\vec x )$ is a coordinate split into time and spatial coordinate. Furthermore
\be h_{mn}(t,\vec x)= U^\dagger (t,0) h_{mn}(\vec x) U (t,0)\,,\ee
where $U$ is the evolution operator and $h_{mn}(\vec x)$ is the canonical coordinate variable. Since $U|0\rangle = |0\rangle$ one obtains
\be G_{mn,rs}(t,\vec x;t',\vec y)=\langle 0|\,h_{mn}(\vec x)U(t,t')h_{rs}(\vec y)|0\rangle \,.\label{canp}\ee

The property $U|0\rangle = |0\rangle$ is a consequence of the fact that the vacuum is a zero-energy\footnote{If the vacuum energy $\e_0$ is not zero, then $U|0\rangle = e^{it\e_0}|0\rangle$.} stationary solution of the Schroedinger equation $(id/dt - H )|\Psi(t)\rangle =0$.
Since LQG is formulated as a canonical quantum gravity theory, and the Schroedinger equation is the analog of the Hamiltonian constraint, then a natural generalization of (\ref{canp}) to the case of LQG will be
\be G_{mn,rs}(t,\vec x;t',\vec y)=\langle \Psi_0|\,h_{mn}(\vec x)U(t,t')h_{rs}(\vec y)|\Psi_0\rangle \,,\label{lqgp}\ee
where $|\Psi_0\rangle$ is a solution of the Hamiltonian constraint which corresponds to the flat spacetime.

The problem with (\ref{lqgp}) is that it is difficult to find the operator $U$ explicitely. Namely, $U$ is a solution of the equation $idU/dt = HU$ where $H$ is the reduced phase space Hamiltonian corresponding to the gauge fixing $T(\vec x)=t$, where $T$ is the time variable, see \cite{mamik}. However, it is difficult to find a well-defined time variable, which is the problem of time in canonical quantum gravity. A way to resolve this problem is to insert in front and behind of the $U$ in the formula (\ref{lqgp}) the identity operator $\sum_s |s\rangle\langle s|$, where $|s\rangle$ are the spin network states. This gives
\be G_{mn,rs}(t,\vec x;t',\vec y)=\sum_{s,s'}\langle \Psi_0|\,h_{mn}(\vec x)|s\rangle\langle s|U(t,t')|s'\rangle\langle s'|\,h_{rs}(\vec y)|\Psi_0\rangle \,.\label{ip}\ee
The matrix elements $\langle s|U(t,t')|s'\rangle$ can be identified with the spin foam amplitude $Z_M (s,s')$ where $M=\S\times [0,1]$ is a four-manifold such that $x=(t,\vec x )\in\S_0$ while $y=(t', \vec y )\in\S_1$ and $\S$ is a three-manifold representing the space while $s\in\S_0$ and $s'\in\S_1$. One can then write
\be G_{mn,rs}(x, y)=\sum_{s,s'}\langle \Psi_0|\,h_{mn}(\vec x)|s\rangle Z_M (s,s')\langle s'|\,h_{rs}(\vec y)|\Psi_0\rangle \,.\label{sfp}\ee

The formula (\ref{sfp}) stays the same even if the vacuum energy $\e_0$ is non-zero, since then $e^{i\e_0 (t-t')}\langle s|U(t,t')|s'\rangle$ can be identified with $Z_M (s,s')$.

The action of the $h$ operator on the spin network states is given by 
\be h_{aa}(\vec x)|s\rangle = \frac{1}{l_P}[j_x (j_x +1) - j_0 (j_0 +1)]|s\rangle \,,\label{ha}\ee
where $j_x$ is the spin of the spin network edge which emanates from the point $x$ in the direction $a$, see \cite{ro2}. Here it is assumed that the point $x$ coincides with a vertex of the spin network $s$ and similarly for the point $y$. The operator $h_{ab}$ is defined as $h_{ab}=E_a^m E_b^n h_{mn}$, where $E_a^m$ is the inverse triad vector density. The $j_0$ is a spin parameter which appears in the flat spacetime vacuum state, and it is proportional to the flat triad vector norm, see \cite{mik2} and section 4. The action of the operator $h_{ab}$ when $a\ne b$ is more complicated, see \cite{ro3}, but the essential features stay the same as in the $a=b$ case. 

By using the formulas (\ref{sfp}) and (\ref{ha}) one can define a propagator $\tilde G_{ab,cd} (x,y)$ such that
\be \tilde G_{aa,bb}(x,y)=\frac{1}{l_P^2}\sum_{s,s'} \bar{\Psi}_0 (s)[j_x (j_x +1) - j_0 (j_0 +1)]Z_M (s, s')[j_y (j_y +1) - j_0 (j_0 +1)]\Psi_0 (s')\,,\label{snp}\ee
where $\Psi_0 (s)=\langle s|\Psi_0 \rangle$. The spatial components of the graviton propagator will be then given by
\be G_{mn,rs}(x,y) =E_m^a (x) E_n^b (x) E_r^c (y) E_s^d (y) \tilde G_{ab,cd}(x,y)\,,\ee
where $E_m^a$ denote the triad one-form densities.

\bigskip
\noindent {\bf 3. Rovelli's propagator}

\bigskip
\noindent In Rovelli's approach \cite{ro2,ro3} the idea is to replace $\bar\Psi_0 (s)\Psi_0 (s')$ in (\ref{snp}) with $\F_0 (s'')$ where $s''$ is a spin network which corresponds to $s \cup s'$. Then $Z_M(s,s')$ is identified with the Hartle-Hawking wavefunction $W(s'')$ in a spin foam model and a natural choice is the Euclidean Barret-Crane spin foam model \cite{bce,pr}. Hence one starts from a slightly diferent definition of the propagator, whose diagonal components are given by
\be \tilde G_{R}(x,y)=\frac{1}{l_P^2}\sum_{s} \F_0 (s)[j_x (j_x +1) - j_0 (j_0 +1)][j_y (j_y +1) - j_0 (j_0 +1)]W (s)\,.\label{rop}\ee

The $W(s)$ is given by
\be W(s)=\int {\cal D}\phi \,O_s (\phi)\, e^{-S_{BC}(\phi)}\,,\ee
where $S_{BC}(\phi)= S_2 (\phi) + \l S_5 (\phi)$ is the group field theory action for the Euclidean BC model \cite{gft}, $S_2$ and $S_5$ are quadratic and quintic forms respectively, $\l$ is the perturbation theory expansion parameter and $O_s (\phi)$ is the group field theory variable representing the spin network state $|s\rangle$ in the Euclidean BC model \cite{mik}. In the lowest-order perturbative expansion of $W(s)$ contribute only the spin networks with five vertices. Since in the BC model field theory only the four-valent spin networks are allowed, one obtains that the lowest-order perturbation theory contribution is from the spin networks $p$ whose graph is the 4-simplex graph. Hence
\be \tilde G_{R}(x,y)=\frac{1}{l_P^2}\sum_{j_l} \F_0 (p)[j_x (j_x +1) - j_0 (j_0 +1)][j_y (j_y +1) - j_0 (j_0 +1)] W_1(p)+O(\l^2)\,,\label{fsnp}\ee
where $j_l$ indicate the spins of the spin network edges and $W_1(p)$ is the BC evaluation of the spin network $p$.

A key ingredient in Rovelli's approach is the choice
\be \F_0 (s) = \exp\left(-\frac{1}{2j_0}\sum_{l,l'} \a_{ll'}(j_l - j_0) (j_{l'} - j_0)+i\sum_l j_l\theta_l\right)\,,\label{vswf}\ee
where $\a$ is a constant matrix and $\th_l$ are parameters to be determined. Since in the BC model context appear only the spin networks whose graphs $\g$ are four-valent, then $\th_l$ can be identified with the dihedral angles associated with the triangulation of $\S$ whose dual one-simplex is the spin network graph $\g$.
 
The Gaussian form of $\F_0 (s)$ implies that the semi-classical limit corresponds to $j_0 \to\infty$, since then  only the spin network states whose edges carry large spins are important. In the limit of large spins one also has
\be W_1(p)\approx \cos\left(\sum_l j_l \theta_l +k_p \pi/4\right) + D(p)\,,\label{tjsa}\ee
where $k_p$ is an integer and $D(p)$ is the contribution from the degenerate configurations.

By combining (\ref{fsnp}), (\ref{vswf}) and that $E_a^m |\F_0\rangle \approx j_0 \d_a^m |\F_0 \rangle$ one obtains
\be G^{(1)}_{R}(x;y)=\frac{1}{j_0^4 N l_P^2}\sum_{j_p}\F (p)[j_x (j_x +1) - j_0 (j_0 +1)][j_y (j_y +1) - j_0 (j_0 +1)] W_1(p)\,, \label{forp}\ee
where $N=\sum_{j_l}\bar\F_0 (p)\F_0 (p)$. In the limit $j_0 \to\infty$ it follows from (\ref{tjsa}) that 
\be G^{(1)}_R (x,y)\approx l_P^{-2} j_0^{-4}{\sum_{j_l}(j_x^2 - j_0^2)(j_y^2-j_0^2)\exp\left(-\frac{1}{j_0}\sum_{l,l'} \a_{ll'}(j_l - j_0) (j_{l'} - j_0)\right)\over \sum_{j_l}\exp\left(-\frac{1}{j_0}\sum_{l,l'} \a_{ll'}(j_l - j_0) (j_{l'} - j_0)\right) }\,,\label{rp}\ee
where the phase factors from (\ref{vswf}) have cancelled the phase factors in (\ref{tjsa}) as well as the $D$ term. The sums over $j_l$ can be converted into the Gaussian integrals which gives
\be G^{(1)}_R (x,y) \approx \const\frac{\a^{-1}_{j_x j_y}}{j_0 l_P^2}\,,\label{asf}\ee
as the leading term. 

Since the flat spacetime state is described by a spin network whose spins are close to $j_0$, then the area $A$ carried by an edge of a 4-symplex spin network $p$ will be approximately $l_P^2 \sqrt{j_0 (j_0 +1)}$. Given that $j_0$ is large, we will have $A\approx l_P^2 j_0$. Let $\s_p$ be the simplical complex whose dual one-complex is the graph of $p$. Then the points $x$ and $y$ are located at the centers of two triangles in $\s_p$. Given that the areas of all triangles in $\s_p$ are approximativelly $l_P^2 j_0$, one can show by using the elementary geometry  that $|x-y|\approx l_P \sqrt{j_0}$. This implies 
\be |x-y|^2 \approx l_P^2 j_0 \,,\ee
so that the formula (\ref{asf}) gives 
\be G_R^{(1)}(x,y)\approx {\textrm{const.}\over |x-y|^{2}}\,,\label{rpa}\ee 
which is the correct large-distance asymptotics given by (\ref{sda}).

\bigskip
\bigskip
\noindent{\bf{4. Nonperturbative propagator}}

\bigskip
\noindent The result (\ref{rpa}) is encouraging but it raises certain questions as far as its range of validity is concerned. It is a perturbative result, so that it is crucial to examine how the higher-orders terms in the perturbative expansion affect it. Another caveat is that the choice of $\F_0 (s)$ given by (\ref{vswf}) is almost certainly not a solution of the Hamiltonian constraint, while the correspondence $ \bar\Psi_0 (s) \Psi_0 (s')\to \F_0 (s'')$ strongly suggests that $\F_0 (s)$ should satisfy the Hamiltonian constraint.

Let us then analyze the formula (\ref{snp}) for $t=t'$, which then corresponds to $Z_M(s,s')=\d(s,s')$. The diagonal components of the propagator will be then given by
\be \tilde G(x;y)=\frac{1}{l_P^2}\sum_{s} \bar{\Psi}_0 (s)[j_x (j_x +1) - j_0 (j_0 +1)][j_y (j_y +1) - j_0 (j_0 +1)]\Psi_0 (s)\,.\label{msnp}\ee
In this case we do not need to know the spin-foam amplitude $Z_M$, and only the knowledge of $\Psi_0(s)$ is required. In \cite{mik2} it was described how to construct $\Psi_0(s)$ in the Euclidean LQG case as a spin-foam state sum for the quantum $SU(2)$ group at a root of unity. This spin network wavefunction is a solution of the Hamiltonian constraint and it dependes on the spin parameter $j_0$ which can be identified as $|E_0|/l_P^2$, where $|E_0|$ is the norm of the flat background triad vector $E_0^a = \int_\D \e_{mnr}E_0^{ma} dx^n \wedge dx^r$, where $\D$ is a triangle of a triangulation of $\S$\footnote{The background triads enter the vacuum wavefunction $\Psi_0 [A]$ through $\exp i\int_\S Tr(A_m E^m_0)d^3 x$. The $\Psi_0 (s)$ is given as the loop transform of $\Psi_0 [A]$ and in order to define the loop transform a triangulation of $\S$ is used, see \cite{mik2}.}. However, at present it is difficult to calculate $\Psi_0 (s)$ due to the lack of efficient tecniques for dealing with sums of products of the corresponding quantum spin network evaluations. 

Let us then assume that exists $\Psi_0(s)$ which is a solution of the Hamiltonian constraint such that its large $j_0$ asymptotics is given by (\ref{vswf}). This is a plausible assumption because the construction for $\Psi_0 (s)$ given in \cite{mik2} has spin network insertions which can be chosen freely, and therefore it is quite possible that a choice of insertions exists such that the large-spin asymptotics is given by (\ref{vswf}). In that case the formula (\ref{msnp}) gives 
\be  G (x,y)\approx l_P^{-2} j_0^{-4}{\sum_{\g}\sum_{j_l}(j_x^2 - j_0^2)(j_y^2-j_0^2)\exp\left(-\frac{1}{j_0}\sum_{l,l'}\a_{ll'}(j_l-j_0)(j_{l'} -j_0)\right)\over \sum_{\g}\sum_{j_l}\exp\left(-\frac{1}{j_0}\sum_{l,l'}\a_{ll'}(j_l-j_0)(j_{l'} -j_0)\right) }\,.\label{npp}\ee
The formula (\ref{npp}) can be obtained by repeating the appropriate steps of the derivation for $G_R^{(1)}$. Note that now the phase factors $\theta_l$ have dissapeared because of the $\bar\Psi_0(s) \Psi_0 (s)$ term in the formula (\ref{msnp}). The sums over the spins in (\ref{npp}) can be then calculated in the same maner as in (\ref{rp}), so that one obtains 
\be G (x,y) \approx \const\frac{1}{j_0 l_P^2}\frac{\sum_{\g}(\a^{-1}(\g))_{j_x ,j_y}\det^{-1/2} \a(\g)}{\sum_{\g}\det^{-1/2} \a(\g)}\,.\label{genp}\ee

In the case when $\a_{ll'}=\a\,\d_{ll'}$ we obtain
\be  G (x,y) \approx \const\frac{\d_{j_x ,j_y}}{j_0 l_P^2}\,,\label{newp}\ee
in the leading order. The leading-order term is non-zero when $j_x =j_y$ and this condition means that the points $x$ and $y$ are at the ends of a spin network edge. One can argue that the distance $|x-y|$ is approximatelly $l_P\sqrt{j_0} $ and the argument is analogous as in the case of the 4-symplex spin network. Namely, if the spin network graph $\g$ is the same as a dual graph of a triangulation of $\S$, then the triangulation is such that all triangles have the area approximatelly $j_0 l_P^2$ and $x$ and $y$ are in the centers of two adjacent triangles and hence $|x-y| \approx l_P \sqrt j_0$. 

If $\g$ is not the same as a dual triangulation graph\footnote{This is a possibility which has to be taken into account because the completnes of the spin network basis requires that the sum in (\ref{npp}) goes over all possible graphs.} of $\S$ then $\g$ can be embedded into a handle decomposition of $\S$ which is a thickening of a dual triangulation graph of $\S$ such that the vertices of $\g$ belong to the zero-handles and the edges of $\g$ belong to the one-handles, see \cite{mik3,marm}. This implies that $x$ and $y$ are at the centers of two adjacent triangles in a subcomplex of the triangulation complex such that the areas of the triangles in the subcomplex are all approximativelly $l_P^2 j_0$. Hence $|x-y| \approx l_P \sqrt j_0$. Therefore the result (\ref{newp}) implies the classical asymptotics (\ref{sda}). 

\bigskip
\bigskip
\noindent{\bf 5. Conclusions}

\bigskip
\noindent Note that if $x$ and $y$ are not on the same edge, then $G(x,y)\approx 0$. This means that we have the correct classical asymptotics for large distances, but for very large distances, i.e. $|x-y|>> \sqrt{j_0} l_P$, the propagator is zero. This is a consequence of the diagonal form of the matrix $\a$, and one can consider such an effect as realistic only if a solution of the Hamiltonian constraint with a diagonal matrix $\a$ can be shown to exists. 

Clearly the properties of the propagator are encoded in the state $\Psi_0 (s)$. Note that the spin network wavefunction can also depend on the intertwiners. In the case of the BC model the intertwiner dependence is not important because the spin network states in that case do not have the intertwiner labels. However, in order to make an exact correspondence with the LQG spin network basis, one must know how the BC spin network states are expressed as linear combinations of the LQG spin network states. 

The construction of $\Psi_0 (s)$ from \cite{mik2} depends explicitely on the intertwiners, and hence it is natural to expect that such a dependence may appear in the large-spin limit. It will be interesting to see whether this happens and how it affects the
graviton propagator asymptotics. Still, it is encouraging that one can find examples of $\Psi_0(s)$ which give the correct classical asymptotics for the propagator and that these examples have the Gaussian form. 

Note that in Rovelli's approach $\Psi_0 (s)$ does not have to be a solution of the Hamiltonian constraint. However, our arguments show that a realistic definition of the propagator requires that $\Psi_0 (s)$ is a solution of the Hamiltonian constraint.
Although the Rovelli ansatz for $\Psi_0 (s)$ is not a likely solution of the Hamiltonian constraint, it can serve as a guide for finding a solution with the required properties, which is quite suitable for the approach of \cite{mik2}. This approach also gives an explanation for the spin parameter $j_0$, which comes from the values of the background triads. 

\bigskip 
\bigskip
\noindent{\bf Acknowledgments}

\bigskip
\noindent
I would like to thank C. Rovelli for discussions. This work has been partially supported by FCT project PCTD/MAT/69635/2006 and ESF project ``Quantum Geometry and Quantum Gravity''.

\end{document}